\documentclass{ws-procs9x6}

\newcommand{\vphi}{\varphi}

\newcommand{\cB}{{\mathcal B}}

\newcommand{\bea}{\begin{eqnarray}}
\newcommand{\eea}{\end{eqnarray}}
\newcommand{\beq}{\begin{equation}}
\newcommand{\eeq}{\end{equation}}

\begin{document}

\title{The Tensor Track: an Update}

\author{Vincent Rivasseau}
\address{Laboratoire de Physique Th\'eorique, CNRS UMR 8627,\\
Universit\'e Paris-Sud, 91405 Orsay, France\\
and Perimeter Institute for Theoretical Physics\\
31 Caroline St. N., ON, N2L 2Y5, Waterloo, Canada\\
$^*$E-mail: rivass@th.u-psud.fr}

\begin{abstract}

The \emph{tensor track} approach to quantum gravity\cite{riv},
is based on a new class of quantum field theories, hereafter
called tensor group field theories (TGFTs) \cite{bgriv,bg1,bgo,bgliv,COR}. 
We provide a brief review of recent  progress and list some desirable properties of TGFTs.
In order to narrow the search for interesting models, we also propose
a set of guidelines for TGFT's loosely 
inspired by the Osterwalder-Schrader axioms of ordinary Euclidean QFT.

\end{abstract}


\section{Introduction}

String theory and loop quantum gravity (LQG), the two leading approaches to quantum gravity, 
are currently stuck by a common problem: the lack of a convincing second-quantized
non-perturbative formulation. 

About twenty years ago, $d$-branes with $d \ge 3$ were recognized as key features of string theory.
The non-perturbative framework that should explain the presence of branes and their beautiful associated dualities 
was called {$M$-theory}, where  $M$ means \emph{matrix}, \emph{m}ystery or \emph{m}agic. But a simple
action for this  $M$-theory is still missing. This problem may be related to the huge and puzzling \emph{landscape} of 
perturbative ordinary string vacua. The way forward may require some radical simplification. 

A candidate for a non-perturbative second quantized formulation of LQG was quite early identified  \cite{RR} as group field theory (GFT) \cite{boulatov,gft1}.
But GFT developed slowly and is not yet the mainstream formulation of LQG; in particular
its correct combinatorics and renormalization
have been found only recently.

The \emph{tensor track} is a generalization of the random matrix 
approach to the quantization of two-dimensional gravity. It rebuilds early tensor models \cite{tensor} and GFT
around new principles derived from the universal properties 
of general random tensors \cite{universal}. It leads to a new class of quantum field theories 
which successfully renormalize GFT divergences, now correctly interpreted as ultraviolet rather than infrared
in the Wilsonian sense\footnote{See \cite{Bahr:2012qj} for renormalization of spin-foams based on lattice-like coarse-graining.}.

Properly supplemented with standard-model matter fields (and possibly supersymmetry?), 
this approach may hopefully some day relate different approaches such as LQG and superstrings through
a framework that we could nickname $T$-theory ($T$ like \emph{tensor} or \emph{t}otal). 
Indeed tensor models contain many embedded matrix models (their \emph{jackets}\cite{GurRyan}). They have therefore
at least in principle the potential to quantize strings and higher dimensional branes on the same footing,
leading to simpler models.

\section{Basic Hypotheses}

In the absence of direct experimental evidence, we expect the search for a good theory of quantum gravity to remain speculative and
based on analogies for quite a while. The tensor track emphasizes quantum field theory, Feynman functional integrals,
phase transitions and the Wilsonian renormalization group. 
Hence it reflects certain prejudices. Other approaches
emphasize other concepts, such as the unification of all interactions, extended symmetries,
canonical quantization, lattice regularizations etc... 

Nevertheless the tensor track is rooted in deep convictions.
Quantum field theory, functional integrals, phase transitions and the Wilsonian renormalization group together 
form our most advanced and most successful tools to understand physical 
systems with \emph{many degrees of freedom}\footnote{We should 
in particular certainly not consider quantum field theory as just a way
to combine special relativity and quantum mechanics, nor renormalization as a way to hide infinities.
They are far more universal, as exemplified by their great success, first of course in particle physics
but also in condensed matter, which is not relativistic,
in statistical mechanics, which is not quantum, etc.}. 
Only quantum field theory together with renormalization can compute accurately (more than ten digits!) 
physical effects 
which involve radiative corrections. 
Only quantum field theory has successfully renormalized all other interactions. 
Although gravity around a flat Minkovski background is not renormalizable in the perturbative sense,
still the most conservative option seems to enlarge quantum field theory in a suitable minimal way
to quantize it.

 The tensor track bets upon the idea that 
quantum gravity should be background independent, even topology-independent, and that classical space-time
and general relativity are effective concepts emerging from a more fundamental  theory 
through one or several \emph{phase transitions},
nicknamed geometrogenesis\cite{gg1}.

Indeed phase transitions, whose modern understanding is provided by the Wilsonian renormalization group, 
are generic features of physical systems with many degrees of freedom. In fact the
myriad of phase transitions leading to composite structures is perhaps the most obvious characteristic of our universe. 
Why would the geometry of the universe itself, with its huge number
of degrees of freedom, not follow this trend?
Another observation is that phase transitions occur at particular scale, hence they could provide an 
explanation for the existence of the Planck and cosmological constant scales in our universe. Finally of course
geometrogenesis nicely fits with the big-bang, as they could be just the same thing.

The main \emph{new} (hence perhaps most controversial) \emph{bet of the tensor track} is to replace 
the ordinary principle of local interactions by an extended notion based on tensor invariance\footnote{
One of the few consensual ideas on the subject is that ordinary locality should
be extended to quantize gravity. Spatial distances or areas smaller than the Planck scale cannot be measured in the usual way, as
the measuring probes would disappear into the black hole created by their own gravitational field.
However this does not mean that meaningful physics necessarily stops at that scale. Transplanckian \emph{scales} could exist in the renormalization group sense even when there is no longer any well-defined notion of  \emph{distance}\cite{riv}; and transplanckian 
\emph{physics} might be detected through indirect but convincing effects,
which may lead to future predictions and physical discoveries.}.
But again this choice is not arbitrary. We believe that quantizing gravity is essentially the same
as correctly randomizing geometry. Since our universe is very large, we need a robust
tool to perform a statistical analysis of \emph{large geometries in three and four dimensions}. The most fundamental tool in probability theory
is the law of large numbers and the central limit theorem. The theory of random matrices and of their invariant interactions provides
the equivalent of this tool to analyze two dimensional geometries. The recently discovered theory of higher rank random tensors and of their invariant interactions\cite{universal} is
the natural candidate for their generalization to dimensions three and four.


\section{GFT's and TGFT's}

Consider a (simple, compact) Lie group $G$, endowed with its natural Haar measure and metric. Complex valued square integrable functions on $G$ form an
associated Hilbert space $H(G)$. The group structure on $G$ allows Fourier analysis. 
$H(G)$ is infinite-dimensional, and admits various approximation schemes through finite $N$-dimensional vector spaces $H_N (G)$.
Rank $d$ tensor fields are defined as elements of the external tensor product of $d$ copies of $H(G)$, or of $H_N (G)$, in which case 
we are interested in letting the cutoff $N$ tend to infinity\footnote{As usual in QFT, tensor fields may be in fact  \emph{distributions} rather than functions but we skip this technicality here.}. The scalar product in $H(G)$ 
or $H_N (G)$  allows to raise and lower tensor indices, hence to contract indices at identical positions between 
a tensor and its complex conjugate.

Equivalently a tensor field can be considered as a function on the product $G^d$ of $d$ copies of $G$, but this erases its tensor aspect.
Group field theory (GFT) nevertheless emphasizes this second point of view; 
it is defined by an action for fields living on  $G^d$\cite{gft1}. 
In the initial example of group field theory, the Boulatov model \cite{boulatov}, the group $G$ is $SO(d)$ or its universal covering group and the field
incorporates a projection which averages over a common group translation of the $d$ variables. This projection trivializes  the holonomies along the faces of any 
Feynman amplitude, hence it implements the $BF$ action on the 2-complex corresponding to the Feynman graph\cite{boulatov}.
However the usual vertex envisioned by ordinary GFT is not a tensor invariant (in the precise sense defined below) and
does not correspond to a stable action. The theory triangulates very singular pseudo-manifolds 
in addition to regular manifolds. No $1/N$ expansion has been found to organize the amplitudes of this theory,
and although many amplitudes become infinite in the no-cutoff limit, they could not be properly \emph{renormalized}.


This situation changed with the discovery of \emph{colored} group field theories\cite{color} 
and of their associated simpler random tensor models \cite{GurRyan}. 
They triangulate better behaved spaces \cite{lost} and admit a 1/N expansion \cite{Gur2} (where $N$ 
is the ultraviolet cutoff).
Their uncolored formulation \cite{universal,BGRuncoloring} rests on the classification 
of all the \emph{$U(N)^{\otimes d}$-invariant interactions} of a pair of complex conjugate random tensors.
It generalizes the standard invariance of (Wishart) matrix models. 
A welcome property is that such interactions are also often \emph{stable} for a suitable sign of the coupling constant, curing one of the main problems of GFT.

The tensor track proposal conjectures that \emph{this}  tensor invariance should be
the proper extension of the ordinary notion of \emph{locality} needed to quantize gravity. 
Rebuilding GFT to incorporate this tensorial aspect of the field. 
we obtain a new class of quantum field theories, namely the TGFTs \cite{bgriv,COR}.

However just as locality in quantum field theory is fundamental but is only an exact property of interactions, not of propagators, 
we expect interesting TGFT's to have non-$U(N)^{\otimes d}$-invariant propagators. It is in
fact the interplay between approximately local propagators and local interactions which launches
the renormalization group flow of coupling constants in quantum field theory, and the same happens in TGFTs.

\section{Desirable Properties}

\subsection{Renormalization}

Just renormalizability is a property shared by all physical interactions except (until now!) gravity. In the renormalization group
sense it is \emph{natural}. Indeed just renormalizable interactions survive long-lived RG flows.
They can be considered the result of a kind of \emph{Darwinian selection} associated to such flows.
Therefore if quantum gravity can be renormalized as proposed in \cite{riva,riv}
it will rely on the same powerful technique that applies successfully to all
other interactions of the standard model. There will be no longer any need for a teleological or 
anthropic interpretation.  

The simplest renormalizable TGFT has been found in dimensions three and four for the $U(1)$ group\cite{bgriv,bgo}. In
dimension 4 it has two unexpected $\phi^6$-like marginal interactions, hence a richer RG flow than 
the usual $\phi^4$ models \cite{bg1}.

\subsection{Asymptotic Freedom} 

Again asymptotic freedom is a property shared by all physical interactions except (until now!) gravity. Indeed
QCD is asymptotically free and the electromagnetic sector inherit at high energy the asymptotic freedom of the unified electroweak theory\footnote{The ultraviolet behavior of the Higgs sector is a subtle issue not considered here.}.

Asymptotic freedom is desirable to build a geometrogenesis scenario for
TGFTs \cite{gg1,gg2,gg3,gg4}, and in fact may be generic in the world of tensors of rank greater or equal to three.
It  has been \emph{already established} for the simplest renormalizable TGFTs in dimension 3 and 4  \cite{bg1,bgo}.
The new locality axiom allows wave function renormalization to compete with coupling constant radiative
corrections, and typically to win in the case of rank $\ge 3$ tensors. Recall
that absence of asymptotic freedom is the rule for the simplest models of scalar, vector and matrix type (except of course non-Abelian gauge theories)
and that asymptotic safety is barely reached for natural matrix field theories such as the Grosse-Wulkenhaar 
model \cite{gwbeta}. In addition, the infrared growth of the coupling constant occurs for the \emph{stable} sign of the interaction,
hence may lead to the discovery of singularities which could represent \emph{unitary matter}. This would improve on single-matrix  
model singularities which lead to (non-unitary) Lee-Yang type singularities.

\subsection{Constructibility}

Constructibility of a quantum field theory means that its perturbative series can be 
uniquely resummed (typically through a kind of Borel resummation)\cite{GJ,Rivbook}.
Physically it is related to \emph{stability and uniqueness of the vacuum}.
It guarantees that at least the perturbative phase of the theory is unique
and mathematically well-defined at small coupling.

TGFTs with stable positive interactions should be constructible, and the corresponding proofs
seem doable, thanks to a new constructive tool
called the {loop vertex expansion} (LVE) \cite{lve}, adapted
to the {extended notions of locality} that govern matrix or tensor models. Significant 
results have been already obtained in this direction\cite{mnrs,universal}. 
We expect the full Borel summability of renormalizable asymptotically free TGFTs
to be more difficult than those of infrared $\phi^4_4$
and of the Gross-Neveu model\cite{Rivbook}, but much simpler than the 
corresponding study for non-Abelian gauge theories.

The existence of such a constructive perspective
is a very important \emph{long term asset} of the tensor track program,
which (to our knowledge) is missing in all other current approaches 
to quantum gravity.

\subsection{Geometricity}

We are ultimately interested in models whose effective infrared physics leads naturally (under suitable boundary conditions) to
our universe, namely a large quasi-flat
four dimensional space-time with a metric obeying the (classical) Einstein equations.

Recently models were developed which incorporate the constraint projector of the $BF$ theory. 
This could lead to a geometrogenesis with a smoother metric. 
The first four-dimensional models of this type have been
proved superrenormalizable on the $U(1)$ group\cite{COR}. We expect $\phi^6$ models in dimension 3 and on the $SU(2)$
group to be just renormalizable \cite{COR2}.

To guide geometrogenesis towards the desired outcome in four dimensions we may have to decorate the most natural
renormalizable TGFT's with additional \emph{geometric conditions}.
Spin foam models, in particular the 4d models which incorporate Plebanski 
simplicity constraints \cite{EPRL} could inspire such decorations. 
We are open to other possibilities, as the only rule is to find 
the \emph{simplest} such models with gravity as their effective limit.

\subsection{Dualities, holography}

Dualities such as Born duality, Langmann-Szabo dualities in the Grosse-Wulkenhaar 
matrix models or the many dualities of string theory
may have interesting analogs in the TGFT world. Such dualities 
could allow integrability and exact solvability of particular TGFT models.
This possibility should be systematically investigated.

Similarly it might be interesting to incorporate some kind of holographic principle in TGFT's. The structure of the boundaries
of TGFT amplitudes, which are themselves lower rank TGFT vacuum amplitudes, suggests some principle of this kind.

\section{Rules for TGFT's}

We sketch now tentative rules for TGFTs that could later evolve into a true axiomatic scheme. Axioms 
embody the long term reflection of the scientific community on the most fundamental aspects of quantum field theory
and are therefore a valuable source of inspiration into unexplored territory such as quantum gravity.  
But at this early stage we intentionally formulated our proposal in a non-technical language. It should not be 
considered rigid nor exclude interesting future developments (for instance Fermionic axioms etc...).
The hard work, which remains entirely to be done, requires a more precise mathematical formulation 
of these rules and the proof that interesting interacting TGFTs indeed obey them. 

According to our conservative analogy-based approach we search for natural analogs of the main 
axioms of Euclidean quantum field theory.
These new rules should imply a new kind of constructive program, for TGFTs. The initial constructive program \cite{GJ,Rivbook} 
is far from complete, as it does not include yet the full construction of the four dimensional Yang-Mills theories. However
a constructive program for TGFTs could actually progress faster in the coming years, since 
interesting asymptotically free models may be free of subtle constructive issues such as Gribov ambiguities which plague the ordinary non-Abelian gauge theories.

Our rules are formulated in terms of approximation schemes based on limits of functional integrals with cutoffs. 
In ordinary quantum field theory we know that each cutoff violates
some axiom; but uniqueness of the limit typically ensures that the 
theory without cutoffs satisfies all of them.

\medskip
\noindent{\bf Rule 1: Tensor Invariance and Positivity of Interactions}
\medskip

This rule replaces locality. The fields are considered both as functions on  $G^d$, the \emph{pre-space}, and as rank-$d$ tensors on $H(G)$. 
The bare functional measure $d\nu  (\phi, \bar \phi)  $ of 
a TGFT should be formally of the form
\begin{equation}
d\nu  (\phi, \bar \phi) = \frac{1}{Z}d\mu_{C} (\phi, \bar \phi) 
 e^{-S_{int}(\phi , \bar \phi)}, \quad  S_{int}(\phi , \bar \phi)= \sum_{b \in \cB} \lambda_{b} I_b (\phi , \bar \phi)\,, 
\end{equation}
where $C$ is the covariance or bare propagator, $\cB$ is a finite set of  \emph{connected positive tensor invariants} 
 labeled by $b$, and the coupling constants
$\lambda_b \in \mathbb{C}, $ should have positive real parts. 
The non formal definition requires as usual to introduce an ultraviolet cutoff $N$, then to control
the limit $N \to\infty$.

\begin{figure}[h]
\begin{center}
\includegraphics[scale=0.4]{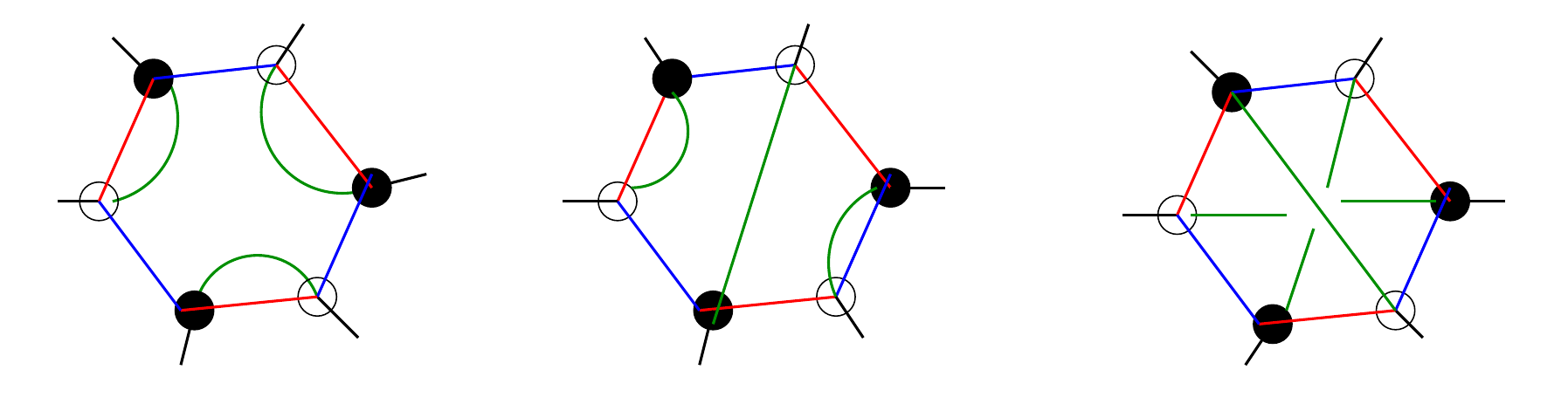}
\caption{Some connected tensor invariants in $d = 3$}
\label{tensinv}
\end{center}
\end{figure}

\emph{Tensor invariants} are obtained by convolution of a set of fields $\vphi$ and $\overline{\vphi}$, in such a way that the $k$-th index of a field $\vphi$ is
always contracted with the $k$-th index of a conjugate field $\overline{\vphi}$, resulting in a polynomial invariant under $U(N)^{\otimes d}$. 
They are canonically represented by closed bipartite \textit{$d$-colored graphs}: each field $\vphi$
(resp. $\overline{\vphi}$) is represented by a white (resp. black dot), and each contraction of a $k$-th index between two fields is pictured as a line with \textit{color} label $k$ linking the two relevant dots (see Figure \ref{tensinv}).  \emph{Connected} invariants correspond to connected graphs. \emph{Positive}
invariants admit a mirror symmetry allowing 
to write them as sums of moduli squares. For instance in Figure \ref{tensinv} the first two tensor invariants admit such a 
symmetry, but not the last one. The conditions on positive real parts for the couplings $\lambda_b $ ensure
stability of the corresponding action.

\noindent{\bf Rule 2: Discrete Permutational Symmetry}
\medskip

We suggest to replace continuous rotational and translational invariance by a discrete permutation invariance,
as appropriate in case of geometric discretizations. Hence we
require that the Schwinger functions should be invariant under the discrete symmetry group 
$\Sigma_d$ with $d!$ elements. In particular tensorial interactions should be symmetrized
over the discrete permutations of the $d$ space-time colors. This implies constraints on the coupling constants:
they should be equal for invariants which differ only by a permutation of colors.

\medskip
\noindent{\bf Rule 3: Clustering}
\medskip

The clustering axiom in ordinary QFT requires the Schwinger functions to decay 
as their external arguments are taken apart. In the pregeometric tensor world 
external arguments of rank $d$ models represent 
boundaries which are themselves colored tensor models of rank $d-1$. There is 
not any good notion of distance yet, hence we should rather ask for a decay 
in the defining parameters of the sum $S_r (c, d_1, \cdots d_c)$ of Feynman amplitudes
which have $r$ external legs defining a boundary with $c$ connected components, each having \emph{degree}
$d_i$.

In case of exponential decay, this would mean that there exist constants $K>0$ and $\epsilon>0$ such that
\beq \vert S_r  (c, d_1, \cdots d_c) \vert  \le K e^{-  \epsilon (r  +  c   + \sum_{i=1}^c d_i )}
\eeq
The number of connected components of the boundary and the sum of their degrees
gives some measure of the complexity both of the topology and of the cellular
structure of that boundary. Decay of this type holds at the perturbative level for the models considered so far\cite{bgriv}.
Beware however that the notion of 
\emph{connectedness} may depend on the models considered\cite{COR}.

\medskip
\noindent{\bf Rule 4: Positivity and Mirror-Positivity of Propagator}
\medskip

The only quadratic invariant for tensors is the mass term which is also local.
Hence \emph{locality coincides with tensor invariance for the 2-point function}. We want to consider
a bare propagator which softly breaks both of these invariances, in order to launch a renormalization group flow. We require that it should have a non-trivial positive spectrum, 
allowing its parametric representation as $C= \int_0^{\infty} e^{- \alpha C^{-1}} d\alpha$. For renormalization group analysis 
to work we also need that it should become approximately local in the ultraviolet regime $\alpha \to 0$.

Most controversially perhaps, we suggest that an analog of Osterwalder-Schrader positivity or 
of the Markov property of ordinary Euclidean fields should also hold for TGFTs. 
Indeed this key property of Euclidean fields allows the
continuation to Lorentzian time and the unitary time evolution of states. Having an analog of that 
in the tensor world could hopefully also leads to a time interpretation of the resulting effective theory.
Without it we could build arbitrarily many convergent quantum field theories in four dimensions, namely non-unitary theories
with \emph{ultraviolet cutoffs}. We certainly want to exclude these. 

Assuming convergence problems solved through constructive theory,
any Euclidean quantum field theory with OS-positive bare propagator and \emph{local interactions} is fully OS positive \cite{GJ}.
Since the interactions we consider are non-local, we have not found yet any analog of this
result for tensors. But we think interesting, at least as a heuristic and tentative rule, to suggest
the bare propagator of our theory to be mirror-positive in the following sense.
The pre-space $G^d$, thanks to the group structure of $G$, comes equipped with $d$ fundamental involutions, 
or mirror symmetries
$$(g_1, ..., g_i, ... g_d) \to S_i (g) = (g_1, ..., g_i^{-1}, ... g_d).$$ 
We could then define a generalized OS-positivity of the propagator $C$, which mean that 
the $p$ by $p$ matrix with matrix elements $C(g_k, S_i (g_l))$ should be positive for any $i\in \{1, \cdots d\}$ 
and any finite collection of $p$ arguments $\{g_k\}$ in $G^d$. This property holds for
propagators admitting a Euclidean K\"allen-Lehmann representation
\beq  C = \int_0^{\infty}  \frac{d\rho (m)}{- \Delta + m},
\eeq
where $\Delta$ is the Laplacian on $G^d$ and $d\rho (m)$ a positive measure. This representation 
excludes better ultraviolet behavior than the one of the Laplacian.

Another strong argument for the Laplacian as propagator in the pre-space comes from the Taylor expansion
around divergent two-point functions required by renormalization\cite{Geloun:2011cy}. Ultimately 
the Laplacian is a natural choice
on $G$, which as a (simple, compact) Lie group comes equipped with 
a differential structure and a metric, so TGFTs should use it.

\section{Conclusion}

Renormalizable 3D and 4D TGFTs  \emph{exist}. The most natural models are asymptotically free,
an encouraging fact for geometrogenesis scenarios.
Adding some pre-geometric content is possible 
at least in some cases (eg $U(1)$ in $D=4$, probably SU(2) in $D=3$).
Axiomatic schemes can be considered for TGFTs,
leading to a new constructive program. Simplified models have been 
proved Borel-summable using the LVE\cite{lve}.

The main open problem is to analyze geometrogenesis of 
natural renormalizable TGFTs and to find the right pre-geometric content 
that would lead to general relativity as effective lower energy physics in dimension 4. 

Success, for probably a long time to come, will be measured in terms of how far
these attempts can be pushed on the mathematical level, how convincingly they imply general relativity as effective
limit and how many applications to different domains they bring. In this last respect the tensor track is 
promising, as it is linked to a growing list of applications of random tensors to statistical mechanics \cite{bonzom1}.


\medskip\noindent
{\bf Acknowledgments}

This paper has greatly benefitted from many discussions with 
J. Ben Geloun, S. Carrozza, R. Gurau and D. Oriti. I would like to thank the organizers of 
the XXIX International Colloquium on Group-Theoretical Methods in Physics
and the Perimeter Institute for support.

{}


\begin{thebibliography}{99}

\bibitem{riv} 
  V.~Rivasseau,
  ``Quantum Gravity and Renormalization: The Tensor Track,''
  AIP Conf.\ Proc.\  {\bf 1444}, 18 (2011)
  [arXiv:1112.5104 [hep-th]].

\bibitem{bgriv} 
  J.~Ben Geloun and V.~Rivasseau,
 ``A Renormalizable 4-Dimensional Tensor Field Theory,''
Commun. Math. Phys. 2012, (DOI) 10.1007/s00220-012-1549-1, 
   arXiv:1111.4997 [hep-th].

\bibitem{bg1} 
  J.~Ben Geloun,
  ``Two and four-loop $\beta$-functions of rank 4 renormalizable tensor field theories,''
  arXiv:1205.5513 [hep-th].

\bibitem{bgo} 
  J.~Ben Geloun and D.~O.~Samary,
  ``3D Tensor Field Theory: Renormalization and One-loop $\beta$-functions,''
  arXiv:1201.0176 [hep-th].

\bibitem{bgliv} 
  J.~B.~Geloun and E.~R.~Livine,
 ``Some classes of renormalizable tensor models,''
  arXiv:1207.0416 [hep-th].

\bibitem{COR} 
  S.~Carrozza, D.~Oriti and V.~Rivasseau,
  ``Renormalization of Tensorial Group Field Theories: Abelian U(1) Models in Four Dimensions,''
  arXiv:1207.6734 [hep-th].

\bibitem{RR} M. Reisenberger, C. Rovelli, Class. Quant. Grav. 18 (2001) 121-140, gr-qc/0002095

\bibitem{boulatov} D. V Boulatov, Mod.Phys.Lett. A{\bf 7}: 1629-1646 (1992).


\bibitem{gft1}
  L.~Freidel,
  ``Group field theory: An overview,''
  Int.\ J.\ Theor.\ Phys.\  {\bf 44}, 1769 (2005)
  [arXiv:hep-th/0505016];
  D.~Oriti,
  ``The microscopic dynamics of quantum space as a group field theory,''
  arXiv:1110.5606 [hep-th].



\bibitem{tensor} M. Gross,  Nucl. Phys. Proc. Suppl. \textbf{25A}, 144-149, (1992); J. Ambjorn, B. Durhuus, T. Jonsson,  Mod. Phys. Lett. \textbf{A6}, 1133-1146, (1991); N. Sasakura, Mod.Phys.Lett. A6
(1991) 2613-2624


\bibitem{universal} 
  R.~Gurau,
  ``Universality for Random Tensors,''
  arXiv:1111.0519 [math.PR].


\bibitem{Bahr:2012qj} 
  B.~Bahr, B.~Dittrich, F.~Hellmann and W.~Kaminski,
  ``Holonomy Spin Foam Models: Definition and Coarse Graining,''
  arXiv:1208.3388 [gr-qc].

\bibitem{GurRyan} 
  R.~Gurau and J.~P.~Ryan,
  ``Colored Tensor Models - a review,''
  SIGMA {\bf 8}, 020 (2012)
  [arXiv:1109.4812 [hep-th]].

  \bibitem{gg1}
  T.~Konopka, F.~Markopoulou and S.~Severini,
 ``Quantum Graphity: a model of emergent locality,''
  Phys.\ Rev.\  D {\bf 77}, 104029 (2008)
  [arXiv:0801.0861 [hep-th]].


\bibitem{color}
 R.~Gurau,
  ``Colored Group Field Theory,''
  Commun.\ Math.\ Phys.\  {\bf 304}, 69 (2011)
  [arXiv:0907.2582 [hep-th]].



\bibitem{lost} 
  R.~Gurau,
 ``Lost in Translation: Topological Singularities in Group Field Theory,''
  Class.\ Quant.\ Grav.\  {\bf 27}, 235023 (2010)
  [arXiv:1006.0714 [hep-th]].
  
\bibitem{Gur2}
  R.~Gurau,
``The 1/N expansion of colored tensor models,''
  Annales Henri Poincar\'e {\bf 12}, 829 (2011)
  [arXiv:1011.2726 [gr-qc]];
R.~Gurau and V.~Rivasseau,
``The 1/N expansion of colored tensor models in arbitrary dimension,''
  Europhys.\ Lett.\  {\bf 95}, 50004 (2011)
  [arXiv:1101.4182 [gr-qc]];
  R.~Gurau,
  ``The complete 1/N expansion of colored tensor models in arbitrary
  dimension,''
  arXiv:1102.5759 [gr-qc].


\bibitem{BGRuncoloring} 
  V.~Bonzom, R.~Gurau and V.~Rivasseau,
 ``Random tensor models in the large N limit: Uncoloring the colored tensor models,''
  Phys.\ Rev.\ D {\bf 85}, 084037 (2012)
  [arXiv:1202.3637 [hep-th]].


\bibitem{riva} 
  V.~Rivasseau,
  ``Towards Renormalizing Group Field Theory,''
  PoS CNCFG {\bf 2010}, 004 (2010)
  [arXiv:1103.1900 [gr-qc]].

  
\bibitem{gg2}
D.~Oriti,
``A quantum field theory of simplicial geometry and the emergence of space-time,''
J.\ Phys.\ Conf.\ Ser.\  {\bf 67}, 012052 (2007)
[arXiv:hep-th/0612301].

\bibitem{gg3}
  D.~Oriti,
``Group field theory as the microscopic description of the quantum space-time
  fluid: a new perspective on the continuum in quantum gravity,''
  arXiv:0710.3276 [gr-qc].

\bibitem{gg4}
  L.~Sindoni,
 ``Emergent models for gravity: an overview,''
  arXiv:1110.0686 [gr-qc].


\bibitem{gwbeta} 
  M.~Disertori, R.~Gurau, J.~Magnen and V.~Rivasseau,
  ``Vanishing of Beta Function of Non Commutative Phi**4(4) Theory to all orders,''
  Phys.\ Lett.\ B {\bf 649}, 95 (2007)
  [hep-th/0612251].
  
  
\bibitem{EPRL} L. Freidel, K. Krasnov, Class. Quant. Grav. \textbf{25}, 125018 (2008) [arXiv: 0708.1595];  J. Engle, R. Pereira, C. Rovelli, Nucl. Phys. B \textbf{798}, 251 (2008), [arXiv:
0708.1236]; J. Engle, E. Livine, R. Pereira, C. Rovelli, Nucl. Phys. B 
\textbf{799}, 136 (2008), [arXiv:0711.0146];   J. Ben Geloun, R. Gurau, V. Rivasseau, Europhys.Lett. 92 (2010) 60008, arXiv:1008.0354 [hep-th]
  A.~Baratin and D.~Oriti,
  Phys.\ Rev.\ D {\bf 85}, 044003 (2012), arXiv:1111.5842 [hep-th].

\bibitem{COR2} 
  S.~Carrozza, D.~Oriti and V.~Rivasseau, work in preparation  
  

\bibitem{GJ}  J.~Glimm, A.~Jaffe,
``Quantum Physics:
A Functional Integral Point of View",
Springer-Verlag, 1987.
  

\bibitem{Rivbook} 
  V.~Rivasseau,
  ``From perturbative to constructive renormalization,''
  Princeton University Press, 1991.

\bibitem{lve} 
  V.~Rivasseau,
  ``Constructive Matrix Theory,''
  JHEP {\bf 0709}, 008 (2007)
  [arXiv:0706.1224 [hep-th]].
  
\bibitem{mnrs} 
  J.~Magnen, K.~Noui, V.~Rivasseau and M.~Smerlak,
  ``Scaling behaviour of three-dimensional group field theory,''
  Class.\ Quant.\ Grav.\  {\bf 26}, 185012 (2009)
  [arXiv:0906.5477 [hep-th]].

\bibitem{Geloun:2011cy} 
  J.~Ben Geloun and V.~Bonzom,
  ``Radiative corrections in the Boulatov-Ooguri tensor model: The 2-point function,''
  Int.\ J.\ Theor.\ Phys.\  {\bf 50}, 2819 (2011)
  [arXiv:1101.4294 [hep-th]].
  
\bibitem{bonzom1}
  V.~Bonzom, R.~Gurau, A.~Riello and V.~Rivasseau,
 ``Critical behavior of colored tensor models in the large N limit,''
  Nucl.\ Phys.\  B {\bf 853}, 174 (2011)
  [arXiv:1105.3122 [hep-th]];
  V.~Bonzom, R.~Gurau and V.~Rivasseau,
  ``The Ising Model on Random Lattices in Arbitrary Dimensions,''
  Phys.\ Lett.\ B {\bf 711}, 88 (2012)
  [arXiv:1108.6269 [hep-th]];
  V.~Bonzom and H.~Erbin,
  ``Coupling of hard dimers to dynamical lattices via random tensors,''
  arXiv:1204.3798 [cond-mat.stat-mech];
 V.~Bonzom, R.~Gurau and M.~Smerlak,
  ``Universality in p-spin glasses",
arXiv:1206.5539.
  
 
\end{thebibliography}
\end{document}